\documentclass[a4paper,11pt]{article}
\usepackage{jcappub}
\usepackage{lineno}
\usepackage{orcidlink}
\usepackage{multirow}
\hypersetup{ linktoc=all,
    colorlinks=true, linkcolor={black},  
       citecolor={blue}, urlcolor={blue}
}

\title{VEGA: Voids idEntification using Genetic Algorithm}

\author[a,1]{Parsa Ghafour\,\orcidlink{0009-0003-2960-1563}\note{corresponding author}}
\author[a]{Saeed Tavasoli\,\orcidlink{0000-0003-0126-8554}}
\author[a]{Mohammad Reza Shojaei\,\orcidlink{0009-0004-7055-2203}}
\affiliation[a]{Department of Astronomy and High Energy Physics, Kharazmi University, 15719-14911, Tehran, Iran}
\emailAdd{P.Ghafour@outlook.com}
\emailAdd{stavasoli@khu.ac.ir}
\emailAdd{mrshojaei@ipm.ir}

\abstract{Cosmic voids are large, nearly empty regions that lie between the web of galaxies, filaments and walls, and are recognized for their extensive applications in the field of cosmology and astrophysics. Despite their significance, a universal definition of voids remains unsettled as various void-finding methods identify different types of voids, each differing in shape and density, based on the method that were used. In this paper, we present VEGA, a novel algorithm for void identification. VEGA utilizes Voronoi tessellation to divide the dataset space into spatial cells and applies the Convex Hull algorithm to estimate the volume of each cell. It then integrates Genetic Algorithm analysis with luminosity density contrast to filter out over-dense cells and retain the remaining ones, referred to as void block cells. These filtered cells form the basis for constructing the final void structures. VEGA operates on a grid of points, which increases the algorithm's spatial accessibility to the dataset and facilitates the identification of seed points around which the algorithm constructs the voids. To evaluate VEGA’s performance, we applied both VEGA and the Aikio–Mähönen method to the same test dataset. We compared the resulting void populations in terms of their luminosity and number density contrast, as well as their morphological features such as sphericity. This comparison demonstrated that the VEGA void-finding method yields reliable results and can be effectively applied to various tracer distributions.}


\begin{document}
\maketitle
\flushbottom
\section{Introduction
\label{sec:intro}}
The large-scale structure of the Universe exhibits an intricate pattern known as the cosmic web, composed of clusters (or knots), filaments, walls (or sheets), and vast low-density regions called cosmic voids (\cite{davis1982survey,peebles2020large}). This structure has been extensively revealed through redshift surveys (e.g., \cite{de1986slice,colless20012df}) and numerical simulations (e.g., \cite{springel2005simulations,vogelsberger2014properties}), which show that galaxies form a network extending across a wide range of spatial scales and redshifts. Within this network, over-dense regions such as filaments, galaxy groups, and clusters concentrate most of the mass, while under-dense regions fill the majority of the cosmic volume \cite{bond1996filaments}. These sparsely populated voids, located between the dense components of the web, are typically bounded by the filamentary structures and contain few or no galaxies \cite{bond1996filaments}.

Since the discovery of voids using Zwicky clusters (\cite{einasto1980structure}) and the first identification of a giant super-void in the Bootes constellation (\cite{kirshner1981million}), numerous observational efforts have expanded our understanding of voids and their role in the cosmic web (e.g., \cite{zeldovich1982giant,geller1989mapping,da1994complete}). These studies have characterized the statistical properties, spatial distribution, and internal structure of voids, offering important insights into the evolution of large-scale structure. Results from both observations and simulations have included measurements of void velocity and density profiles (\cite{paz2013clues,hamaus2016constraints}), analyses of their auto-correlation function and clustering bias (\cite{clampitt2016clustering}), and investigations into their depth, abundance, and geometric properties (\cite{sutter2012public,nadathur2019beyond}).

Voids have emerged as powerful cosmological laboratories in both simulations and observations to extract information about the nature and evolution of the Universe (e.g., \cite{pisani2015counting,hamaus2016constraints,ade2016planck}). Because they lie close to the linear regime (e.g., \cite{hamaus2014universal}), their relatively simple dynamics make them well-suited for cosmological analysis (\cite{aragon2013hierarchical,tavasoli2021void}). They have been used to constrain key parameters such as the matter density and dark energy density (\cite{dekel1993omega, bernardeau1997omega,fliche2010lambda,pisani2015counting,bos2012darkness}), and to probe phenomena including the presence of magnetic fields in low-density regions (\cite{beck2013magnetic,taylor2011extragalactic}). Studies have also employed void observations to interpret anomalies in the Cosmic Microwave Background (CMB), such as the cold spot (\cite{rudnick2007extragalactic}) and large-angle anisotropies (\cite{inoue2006local}), and to explore alternative explanations for the accelerating expansion of the Universe (\cite{moffat2005gravitational,celerier2007accelerated,alexander2009local}). In addition, cosmic voids have been exploited to test the standard cosmological model (e.g., \cite{hamaus2020precision,aubert2022completed,woodfinden2022measurements}), and forecasts suggest that upcoming redshift surveys will significantly enhance the constraining power of void statistics (\cite{contarini2022euclid,bonici2023euclid}). Voids serve as sensitive probes of dark energy and modified gravity (\cite{van2014voids,pisani2015counting,pisani2019cosmic}), with their size function and density profiles offering direct tests of these models (\cite{voivodic2017modeling,perico2019cosmic}). Their cosmological utility encompasses a variety of measurable properties, such as density and velocity profiles (\cite{tavasoli2021void}), redshift evolution (e.g., \cite{nadathur2020completed,aubert2022completed}), and gravitational lensing signals that reveal de-magnification effects as photons traverse under-dense regions (\cite{camacho2024cosmic}). Various studies have probed the integrated Sachs–Wolfe (ISW) effect (\cite{sachs1967perturbations}) by stacking temperature anisotropy maps on void positions (e.g., \cite{ilic2013detecting}), and also explored void signals in Compton y-maps to study the thermal Sunyaev–Zeldovich (tSZ) effect (\cite{alonso2018measurement,li2024cross}). Statistical analyses of void size distributions have been used to estimate key parameters such as $S_{8}$ and $H_{0}$ (e.g., \cite{contarini2024perspective}), and voids have also proven valuable in constraining the sum of neutrino masses by examining how neutrino properties influence void sizes and distributions (e.g., \cite{kreisch2019massive,contarini2021cosmic}). The dynamical structure within voids affects cosmic flow patterns, influencing the peculiar motions of galaxies near walls and filaments (\cite{van2016zeldovich,valles2021void,bermejo2024topological}). Furthermore, the quasi-spherical shape of voids makes them well-suited for Alcock–Paczynski tests (\cite{alcock1979evolution,nadathur2019beyond,hamaus2022euclid}), and their theoretical size distribution, known as the void size function(\cite{verza2019void}), continues to be a powerful probe of dark energy (e.g., \cite{contarini2023cosmological,song2024cosmological}).

Due to their importance, cosmic voids are likely to be at the forefront of cosmological research. However, the absence of a standard definition poses a significant challenge, and a consensus on how to define a void has yet to be established.

Various methods have been developed for identifying cosmic voids. The primary objective of automating this process is to obtain a consistent representation of void structures across different tracer (or galaxy) distributions. One of the main challenges, however, lies in clearly defining the characteristics of the structures being sought.

Some void finder algorithms consider voids as spheres, or a finite number of spheres or other topological shapes \cite{kauffmann1991voids,muller2000voids,hoyle2002voids,colberg2005voids}. These algorithms can be classified based on their assumptions about the shape of the identified voids. They differ in their focus on either completely empty regions or under-dense areas. For instance, certain algorithms approximate the minimum diameters of voids by locating empty spheres within tracer distributions \cite{einasto1989structure}, while others extend this framework to identify ellipsoidal empty regions with varying axes and orientations \cite{ryden1995measuring,ryden1995voids}, or attempt to grow voids from an initial cube by adding neighboring empty regions \cite{kauffmann1991voids}. Some methods construct voids as unions of overlapping sub-regions under integrated density constraints (e.g., \cite{paz2023guess}), and several leverage geometrical decompositions of space through tessellation and evaluate local density gradients to delineate void basins (e.g., \cite{neyrinck2008zobov,platen2007cosmic,nadathur2019revolver,sutter2015vide}). It’s worth noting that, like other techniques that avoid shape assumptions, VEGA does not enforce any specific geometric constraints on the structure of the voids it identifies.

Density estimation plays a central role in void detection, and a variety of methods have been developed for this purpose. Simpler techniques include nearest-neighbor estimators \cite{aikio1998simple} and kernel-based smoothing \cite{shandarin1989large}, which provide local density estimates based on discrete tracer distributions. More advanced approaches rely on adaptive interpolation frameworks, such as those based on tessellation \cite{lavaux2010precision,neyrinck2008zobov}, or apply grid-based thresholding schemes to identify underdense regions. Some methods go further by utilizing phase-space information to pinpoint regions that have not undergone shell-crossing, allowing for a dynamical distinction between void and non-void environments \cite{falck2012origami}.

These definitions have some theoretical justification, as under-dense regions expanding within a homogeneous background tend to become more spherical over time \cite{del2007dark}. They are also geometrically straightforward and do not require complex algorithms. However, the actual Universe contains many voids that are often more polyhedral than spherical (e.g., \cite{icke1987fragmenting}), or exhibit even more general shapes \cite{shandarin2006shapes}.

VEGA represents the first application of Genetic Algorithm (GA) analysis for identifying and locating void block cells. GA is a widely used method for generating high-quality solutions to both constrained and unconstrained optimization and search problems \cite{mitchell1998introduction}. It is an adaptive heuristic search technique that belongs to the broader class of evolutionary algorithms \cite{petrowski2017evolutionary}. Inspired by natural selection and genetics, which are the processes that drive biological evolution, these algorithms intelligently guide random searches by using historical performance data to explore regions of the parameter space with higher potential.

The Genetic Algorithm iteratively modifies a population of individual solutions. Each generation consists of a set of individuals, where each individual represents a point in the parameter space and a potential solution. During each generation transition, the algorithm randomly selects individuals from those that have survived the current population, specifically those with fitness values above a defined survival threshold, using its selection operator to serve as parents. These parents are then combined through crossover and mutation operations to generate offspring for the next generation (e.g., \cite{srinivas1994adaptive}). Over successive generations, the population evolves toward optimal solutions.

Genetic Algorithms offer a flexible and robust framework for optimization problems that involve complex, high-dimensional, non-differentiable parameter spaces, and their population-based search strategy combined with stochastic variation through mutation and crossover enables efficient exploration of diverse solution landscapes (e.g., \cite{burkhart2023neuroevolutionary}). These features make Genetic Algorithms effective for multi-objective problems and support a bias-free search strategy (\cite{nesseris2012new}), particularly when integrating multiple datasets and navigating large parameter spaces with minimal assumptions (\cite{nayak2022application}). Their adaptability allows for the integration of multiple constraints (\cite{medel2023cosmological}), which could be highly valuable in cosmological analyses involving structure detection and classification. In VEGA, this flexibility is utilized to filter out over-dense regions by incorporating three metrics: background volume fraction, mean nearest-neighbor distance, and luminosity density contrast, as part of the void identification process.

These algorithmic strengths have motivated a wide range of applications of Genetic Algorithms in cosmology and astrophysics, where complex parameter spaces and model-independent reconstructions are often required. For instance, a method was developed to reliably determine the orbital parameters of interacting galaxies, which has been applied to both artificial and real data \cite{wahde2001determination}. An intriguing variant of the canonical Genetic Algorithm has been successfully employed to tackle various problems, including the challenging task of finding the orbital parameters of planets orbiting 55 Cancri, based on radial velocity measurements from that stellar system \cite{canto2009simple}. GA techniques have also been used to reconstruct the expansion history of the universe in a model-independent manner \cite{nesseris2010model}, and to analyze Type Ia supernova data to extract model-independent constraints on the evolution of the dark energy equation of state \cite{bogdanos2009genetic}.

In a broader multi-probe context, Genetic Algorithms have been applied to combine information from several cosmological observations, including supernovae, Baryon Acoustic Oscillations (BAO), and the growth rate of matter perturbations \cite{nesseris2012new}. A complementary application focuses on late-time cosmological tensions, using low-redshift background and redshift-space distortion data to identify trends through GA-based reconstruction \cite{gangopadhyay2023phantom}. Another study introduced a methodology for exploring local features in the primordial power spectrum by coupling a Genetic Algorithm pipeline with a Boltzmann solver and CMB data \cite{lodha2024searching}. Similarly, a GA framework has been used to reconstruct the CMB temperature anisotropy map on large angular scales through an internal linear combination (ILC) of final-year WMAP and Planck observations \cite{nayak2022application}. A related application in particle astrophysics involves the use of Genetic Algorithms to optimize parameters governing cosmic ray injection and propagation models \cite{luo2020genetic}.

GA-based non-parametric techniques have also been developed for reconstructing projected lensing mass distributions in strongly lensed systems \cite{liesenborgs2006genetic}, and radiative transfer codes have been combined with GA to automate dust spectrum fitting for AGB stars \cite{baier2010fitting}. Beyond these applications, a parallelized GA has served as the foundation for an autonomous fitter of spectra from massive stars with stellar winds \cite{mokiem2005spectral}, and a highly parallelized and distributed GA has been implemented to determine the globally optimal parameters of stellar models \cite{metcalfe2000genetic}. Additionally, a robust approach has been proposed to optimize telescope scheduling using GA for identifying Pareto-optimal solutions \cite{kubanek2010genetic}.

This paper is organized as follows. Section \ref{sec:data} describes the input dataset, which includes the distribution of tracers (galaxies) used in the analysis. Section \ref{sec:mtd} introduces the new method proposed for identifying cosmic voids. Section \ref{sec:res} presents the analysis and evaluation of the outcomes obtained by applying both the VEGA and Aikio-Mähönen (AM) \cite{aikio1998simple} methods to an identical tracer (galaxy) distribution. It also compares the final void characteristics identified by VEGA, highlighting parameter effects and their contrast with results from the AM method. Ultimately, Section \ref{sec:sum} presents a brief summary of the study.

\section{Data: Millennium
\label{sec:data}}
A semi-analytical model (SAM) is a phenomenological framework that uses simplified equations to describe the key baryonic physical processes involved in galaxy formation and evolution. Early SAMs, which were combined with merger trees derived from analytical approaches such as the Press–Schechter formalism \cite{press1974formation} and its extensions \cite{bond1991excursion,sheth2001ellipsoidal}, successfully reproduced galaxy populations with properties in close agreement with observational data \cite{kauffmann1993formation,kauffmann1999clustering,somerville1999semi,cole2000hierarchical}. Advances in computational methods and the development of large dark matter-only (DMO) simulations, such as the Millennium Simulation \cite{springel2005simulations}, enabled the next generation of SAMs to operate on halo merger trees extracted from these simulations \cite{croton2006many,de2006formation}. Today, most SAMs utilize simulation-based merger trees and incorporate a broad range of physical processes, including gas cooling, disk and bulge formation, stellar and black hole feedback, and environmental effects \cite{guo2013galaxy,croton2016semi,lacey2016unified,cora2018semi,lagos2018shark,henriques2020galaxies}.

L-GALAXIES is a semi-analytical model (SAM) of galaxy formation that is typically implemented on sub-halo merger trees derived from the Millennium \cite{springel2005simulations} and Millennium-II \cite{boylan2009resolving} N-body simulations. Both simulations assume a $\Lambda$CDM cosmology, with parameters obtained from a combined analysis of the 2dFGRS \cite{colless20012df} and the first-year WMAP data \cite{spergel2003first}: $\Omega_{M} = 0.25$, $\Omega_{b} = 0.045$, $\Omega_{\Lambda} = 0.75$, $n_{s} = 1.0$, $\sigma_{8} = 0.9$ and $H_{0} = 100 h$ $(km/s/Mpc)$ with $h = 0.73$. For the dataset used in this work, as outlined in \cite{ayromlou2021galaxy}, the original cosmology has been rescaled using the method of \cite{angulo2010one}, as updated by \cite{angulo2015cosmological}, to match the best-fitting cosmological parameters derived from the first-year Planck data. The underlying cosmology of the dark matter simulations and thus the galaxy formation model, follows the values for matter, baryon, and cosmological constant density parameters, $\Omega_{M} = 0.315$, $\Omega_{b} = 0.0487$, and $\Omega_{\Lambda} = 0.685$ respectively, and $n_{s} = 0.96$ for the scalar spectral index, $\sigma_{8} = 0.829$ for the fluctuations amplitude, and $h = 0.673$ for the Hubble constant.

In this study, we selected all simulated galaxies from the L-GALAXIES catalog that are brighter than approximately $-18$ in the $r$-band filter, have stellar masses exceeding $10^{8}\,M_{\odot}$, and reside at a redshift of $z = 0$ within a volume of $300 \times 300 \times 300\,\text{Mpc}^{3}$. Our final sample comprises approximately 278000 galaxies that meet the selection criteria outlined above, enabling us to focus on a range of galaxies that are crucial for understanding the structure formation and identification of under-dense and over-dense regions.

\section{Method
\label{sec:mtd}}
The VEGA method consists of six phases that are applied sequentially to the input dataset. These phases involve:

(\ref{sec:mtd_grd}) Adding a grid of points to the input dataset,

(\ref{sec:mtd_VT}) Voronoi partitioning and convex hull volume calculation,

(\ref{sec:mtd_LD}) Luminosity density calculation of Voronoi cells,

(\ref{sec:mtd_GA}) Genetic Algorithm analysis for over-density filtering,

(\ref{sec:mtd_fsv}) Using cell centers to identify and order the seed points,

(\ref{sec:mtd_fsv2}) Constructing the final structure of voids.

\subsection{Grid Points
\label{sec:mtd_grd}}
In the initial phase, VEGA inserts a grid of points into the dataset, with each point spaced $d_G$ $Mpc$ apart from its horizontal and vertical neighbors. These grid points are treated as original data points throughout all phases of the algorithm and are considered as galaxies with zero luminosity. Their primary role is to enhance the execution of the Voronoi tessellation (Section \ref{sec:mtd_VT}), thereby improving coverage and accessibility within the dataset space.

\begin{figure*}[t]
\centering
\includegraphics[width=0.62\textwidth]{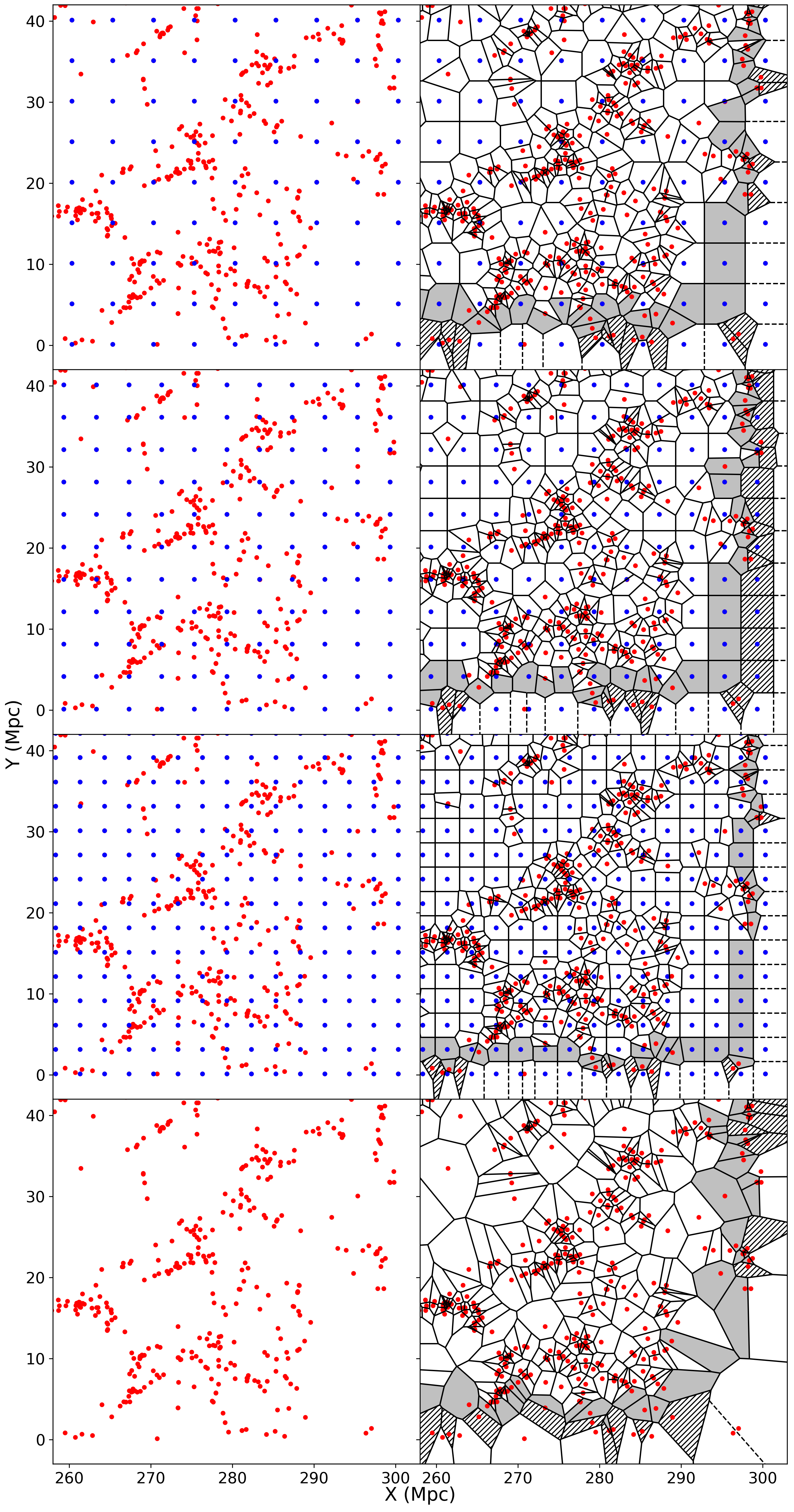}
\caption{
The effect of adding grid points with different $d_G$ values is illustrated. The rows from top to bottom correspond to $d_G$ values of 5, 4, 3, and no grids respectively. \textit{(Left Column)} A slice of the projected galaxy distribution is shown in red, with the added grid points displayed in blue. \textit{(Right Column)} The impact of using grid points alongside galaxies on the Voronoi diagrams is shown. Hatched cells are those with one or more vertices located outside the dataset boundaries and are excluded from further analysis. Cells colored in gray represent marginal cells that are adjacent to the dataset boundary.}
\label{fig:1}
\end{figure*}

Adding these grid points offers several advantages. First, it modifies the shapes of Voronoi cells by reducing the prevalence of large cells corresponding to a data point and removing the irregularly shaped cells with sharp edges, leading to a better shape of the final voids without sharp or irregular shapes and edges. Second, it increases the overall number of Voronoi cells, particularly under-dense ones that are more likely to be associated with void regions, thereby enhancing spatial accessibility. Third, it improves boundary conditions by minimizing sharp-edged, oddly shaped cells near the edges of the dataset and refining the shape of marginal cells, which helps reduce loss of usable space in peripheral regions. The impact of incorporating grid points with various $d_G$ values is illustrated in the right column of Figure \ref{fig:1}.

\subsection{Voronoi Cells and Volumes
\label{sec:mtd_VT}}
The second phase consists of three sub-phases. First, VEGA employs the Voronoi tessellation technique \cite{aurenhammer1991voronoi} to partition the dataset space. To achieve this, the Voronoi diagram of the tracer distribution needed to be constructed. A Voronoi diagram divides the dataset space into cells surrounding a specific set of objects. In our scenario, these objects correspond to the tracers (galaxies and grid points) within the distribution. Each tracer is associated with a distinct cell, known as a Voronoi cell, encompassing all volume of the dataset space closer to that tracer than any other. It is important to mention that VEGA utilizes the standard Euclidean metric for distance measurements.

Once the Voronoi cells and their corresponding vertices have been identified, it is necessary to exclude certain cells. Initially, VEGA eliminates all cells that contain one or more vertices at infinity, as their volume would become infinite when calculating the volume of each cell (section \ref{sec:mtd_VT}). These vertices are illustrated with dashed lines in the right column of figure \ref{fig:1}. Then, VEGA removes those cells that have any vertices located outside the dataset boundaries. This is due to the large and irregular shapes of some of these cells, which could adversely impact the implementation of the Genetic Algorithm in identifying void block cells (section \ref{sec:mtd_GA}). These cells are hatched in the right column of figure \ref{fig:1}. Finally, those cells adjacent to eliminated ones are marked as marginal cells, which are used later to identify and exclude marginal voids during the final phase of the void identification process (Section~\ref{sec:mtd_fsv2}). These marginal cells are shown in gray in figure \ref{fig:1}.

Then, to calculate the volume of each cell, VEGA employs the Quickhull algorithm \cite{barber1996quickhull} to construct the Minimum Convex Polygon (MCP) or convex hull around each cell \cite{virtanen2020scipy}. This construction utilizes the coordinates of the vertices identified through the Voronoi tessellation technique in the previous phase.

The convex hull of a set of points is defined as the smallest convex set that encompasses all the points \cite{fan1984some}. In three-dimensional space, this can be visualized as the shape formed by stretching an elastic band around the outermost points of the set, thereby enclosing all internal points \cite{sack1999handbook}. To construct the convex hull of a cell, it is first necessary to compute the convex set of its vertices determined by Voronoi tessellation technique. For this purpose, VEGA utilizes the Quickhull algorithm \cite{virtanen2020scipy,barber1996quickhull}. The convex hull of the cell is then defined as the minimal convex polyhedron in which every vertex of the cell lies on the boundary of this polyhedron.

Once the convex hull of a cell is constructed, its volume is calculated by decomposing the convex polyhedron into tetrahedrons. The algorithm \cite{virtanen2020scipy} selects a random interior point and generates tetrahedrons by connecting it to the vertices of each face. The summation of these volumes yields the total cell volume $V$ \cite{cohen1979two}.

\subsection{Luminosity Density
\label{sec:mtd_LD}}
In the third phase, VEGA computes the luminosity density of each cell. Given that the dataset is now comprises of two types of points (galaxies and grid points), cells associated with grid points, which contain no galaxies, will have a luminosity density of zero. However, for cells corresponding to galaxies, VEGA calculates the luminosity density by:
\begin{equation}
\label{eq:3.1}
\rho_{L_{i}} = \frac{L_{i}}{V_{i}} \,,
\end{equation}
where $L_{i}$ represents the luminosity of the corresponding galaxy and $V_{i}$ denotes the volume of the associated cell (section \ref{sec:mtd_VT}).

Regarding the background luminosity density ($\rho_{L_b}$), only cells with luminosity density greater than zero that correspond to galaxies are considered in the calculation. This method naturally accounts for variations introduced by the grid spacing parameter $d_G$. As $d_G$ decreases, the cell volumes become smaller, resulting in higher luminosity densities for both individual cells and the background. This type of calculation ensures that the effect of using grid points influences not only the luminosity density of the cells but also the background luminosity density. In this way, as $d_G$ decreases, the background density increases alongside the cell densities.

A traditional background definition considers the total volume of the background rather than only the cells associated with galaxies. As a result, it remains constant regardless of grid spacing. In this case, decreasing $d_G$ would increase the contrast, making cells appear more over-dense and potentially distorting the void identification. By allowing the background to scale with the grid spacing, this formulation avoids such bias.

\subsection{Genetic Algorithm Analysis
\label{sec:mtd_GA}}
In the fourth phase, VEGA builds on the results from the previous phases and applies Genetic Algorithm (GA) analysis \cite{mitchell1998introduction} to identify and determine which cells are classified as void block cells, which form a continuous and under-dense substructure of the background, serving as the building blocks of voids.

To identify the void block cells, VEGA applies a Genetic Algorithm (GA) analysis. This approach is designed to filter out over-dense regions while preserving a continuous under-dense volume of the background. The GA operates on possible combinations of Voronoi cells (section \ref{sec:mtd_VT}), encoded in chromosomes as potential solutions. Each of these chromosomes contains a sequence of binary genes, where each gene represents the inclusion (1) or exclusion (0) of a specific cell, corresponding to galaxies or grid points. However, all cells corresponding to grid points, which have a luminosity density of zero, are permanently included and remain fixed in all chromosomes. These cells represent the empty regions of the dataset and are more likely to be part of a void. The optimization is carried out over the remaining cells with non-zero luminosity density, which correspond to galaxies.

The GA begins by initializing a population of chromosomes, with all zero-luminosity (grid) cells included by default. The remaining cells, which have luminosity density greater than zero, are randomly initialized. The total number of chromosomes in each generation is determined by the expression:
\begin{equation}
	\label{eq:3.4}
	N_{ch} = \epsilon\times\left\lfloor\frac{n^{gc}}{1000}\right\rfloor \,,
\end{equation}
where $n^{gc}$ is the number of galaxy-associated cells, and $\epsilon$ regulates the population size. The effect of different values of $\epsilon$ is presented in the next section (\ref{sec:res}). The correlation between the number of chromosomes $N_{ch}$ and the total number of galaxy-associated cells $n^{gc}$ arises from the stochastic nature of the Genetic Algorithm. Since the inclusion or exclusion of cells in the initial population is determined randomly, the analysis of a larger number of galaxies benefits from an expanded chromosome pool, which improves the algorithm’s ability to sample and explore the parameter space effectively. To ensure robust performance, a sufficient set of potential solutions (chromosomes) is required to create a diverse and representative parameter space, allowing the algorithm to progress toward better scores across generations until it converges on the optimal solution. Furthermore, due to the influence of the mutation operator, a larger chromosome pool enables the algorithm to suppress unsuitable mutations and preserve beneficial ones.

After the creation of the initial population, chromosomes are scored based on their position in the three-dimensional parameter space comprising the background volume fraction ($V_b$), the mean nearest-neighbor distance ($\bar{d}_{nn}$) of the galaxies, and the luminosity density contrast ($\delta_{L_{ch}}$). For each chromosome, the background volume fraction $V_b$ is calculated by dividing the total volume of the selected cells (those for which the corresponding gene has a value of 1 and is considered included) by the total volume of all cells. The mean nearest-neighbor distance $\bar{d}_{nn}$ is obtained by considering only those galaxies for which the corresponding cell are selected; for each such galaxy, the distance to its nearest neighboring galaxy is measured, and the average of these distances gives the $\bar{d}_{nn}$ value for the chromosome. The luminosity density contrast is defined as:
\begin{equation}
\label{eq:3.2}
\delta_{L_{ch}} = \frac{\rho_{L_{ch}}-\rho_{L_{b}}}{\rho_{L_{b}}} \,,
\end{equation}
where $\rho_{L_b}$ is the background luminosity density, and $\rho_{L_{ch}}$ is the total luminosity density of the selected cells in each chromosome and is calculated by:
\begin{equation}
	\label{eq:rho_ch}
	\rho_{L_{ch}} = \frac{L_{ch}}{V_{ch}}
\end{equation}
where $L_{ch}$ is the total luminosity of the selected cells that correspond to galaxies (since grid-point cells have zero luminosity), and $V_{ch}$ is the total volume of all selected cells corresponding to both galaxies and grid points. Then, the chromosomes are scored ($\zeta_{ch}$) by computing the Euclidean distance from the origin of this parameter space:
\begin{equation}
\label{eq:3.3}
\zeta_{ch} = \sqrt[]{{V_b}^2+\bar{d}_{nn}}^2+{\delta_{L_{ch}}}^2 \,,
\end{equation}

Next, since there is no prior generation before the initial population to establish a survival threshold, the top half of the chromosomes are retained as survivors to seed the algorithm, while the remaining ones are discarded. The second-generation chromosomes are then produced through a combination of single-point crossover and mutation operators \cite{srinivas1994adaptive}. In each single-point crossover operation, two parent chromosomes are randomly selected from the survivors, and their gene sequences are exchanged at a randomly chosen point to form two new offspring, which are included in the second generation. The mutation operator subsequently introduces small stochastic changes to these chromosomes. For each gene corresponding to a cell associated with a galaxy, a random number within the range 0 to 1 is generated, and if this number is less than the mutation rate $r_\mu$, the gene is flipped. The mutation operator does not modify genes associated with grid points, as they remain fixed across all chromosomes. The mutation rate $r_\mu$, along with $\epsilon$, serves as one of the tunable parameters of the GA. The effect of the mutation rate is also explored in the next section (\ref{sec:res}).
	
Then, the second-generation chromosomes are scored (Eq. \ref{eq:3.3}). In this stage, the survival threshold is set by the highest-scoring chromosome of the previous generation, and those with scores higher than this threshold are retained as survivors, while the remaining ones are discarded.

The GA then progresses through successive generations. In each generation, chromosomes are scored, survivors are retained, and the next generation is created using single-point crossover and mutation. The algorithm proceeds through generations until either no chromosome achieves a score higher than the best chromosome of the previous generation, or no surviving pair of parents remains to carry forward. If there is no chromosome with a higher score, the best chromosome from the previous generation is retained as the final result. If there is only one chromosome with a higher score, then that chromosome is considered as the final result.

As the Genetic Algorithm evolves the chromosomes through successive generations, it favors solutions that exhibit desirable characteristics across all three dimensions of the parameter space. Chromosomes that encompass cells located in less dense regions naturally contribute to a larger background volume. Those that include galaxies with greater mean distances to their nearest neighbors tend to reflect lower local galaxy densities. Additionally, the luminosity density contrast plays a key role in shaping the final selection. While the scoring formula (Eq. \ref{eq:3.3}) considers only the magnitude of density contrast, the influence of other two parameters leads the chromosomes toward more negative density contrast values through successive generations. In underdense regions, cells corresponding to galaxies have larger cell volumes, greater distances to their nearest neighbors, and more negative luminosity density contrasts than cells in dense regions. These characteristics contribute to a larger background volume fraction, a larger mean nearest-neighbor distance, and a greater magnitude of the luminosity density contrast. By contrast, cells corresponding to galaxies in dense regions, although they may exhibit a high magnitude of luminosity density contrast and could otherwise be favorable for the algorithm, are suppressed due to the influence of the other two parameters. Such cells have both significantly smaller volumes and shorter nearest-neighbor distances, which together reduce the chromosome’s overall score.

\begin{figure*}[t]
\centering
\includegraphics[width=0.92\textwidth]{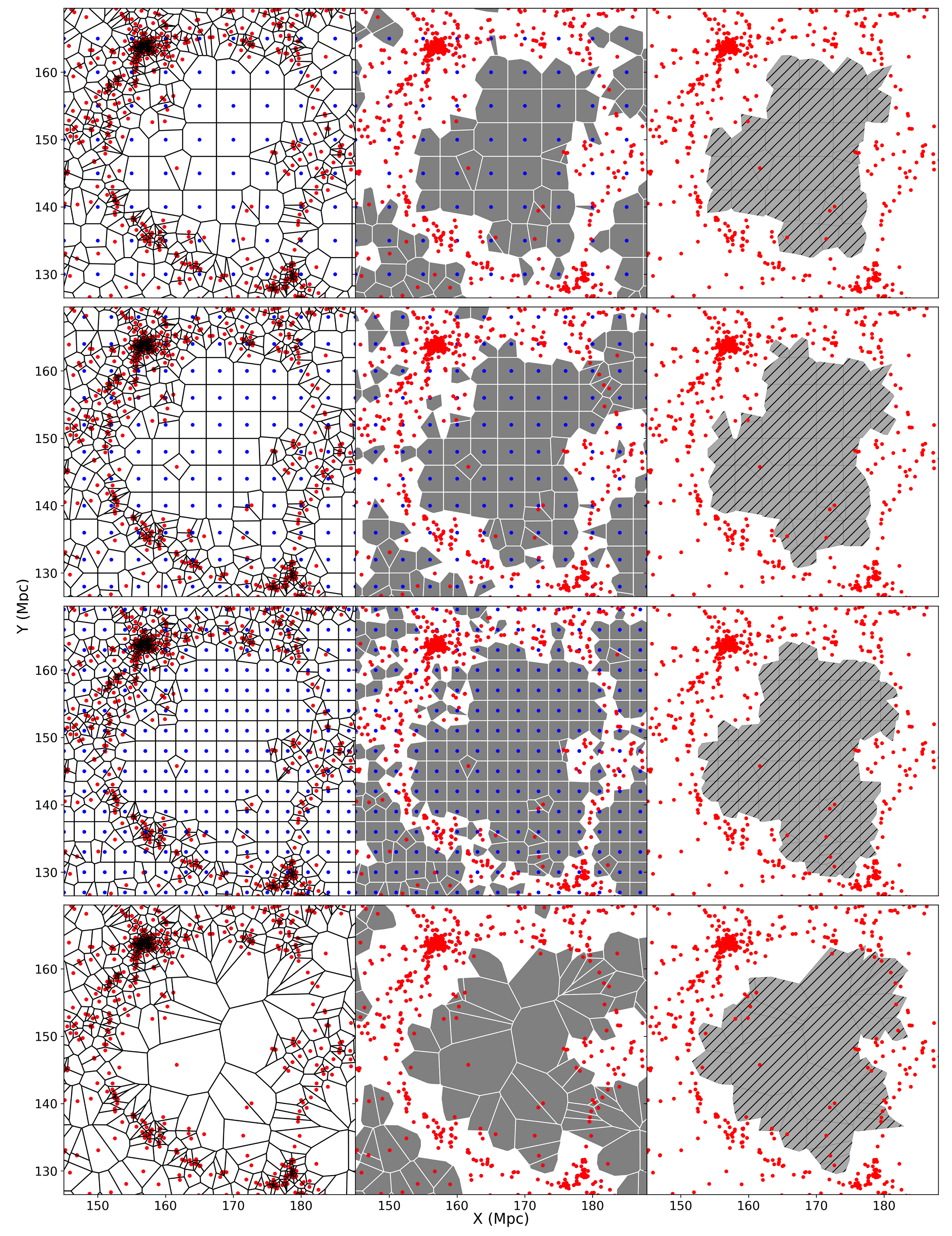}
\caption{
Results of the Genetic Algorithm analysis are shown, with the rows from top to bottom corresponding to $d_{G}$ values of 5, 4, 3, and to the case without grids, respectively. \textit{(Left)} column displays the Voronoi cells associated with galaxies and grid points. \textit{(Middle)} column presents the GA results after post-processing corrections, where cells identified as void block cells are colored in grey. \textit{(Right)} column illustrates the final void identified by VEGA, with the corresponding cells colored in grey. Galaxies are shown in red, and grid points in blue.}
\label{fig:2}
\end{figure*}

After obtaining the final result of the GA analysis, VEGA applies two correction steps to refine the final void block cells. First, any cell marked as included but lacking neighboring included cells is removed to eliminate isolated fragments. Second, any cell not initially included but surrounded entirely by included neighbors is added. This correction ensures that no cells remain enclosed within a void. The corrected GA result, representing the final set of void block cells, is depicted for different grid spacings in Figure~\ref{fig:2}.

\subsection{Seed Points
\label{sec:mtd_fsv}}

In this phase, VEGA follows a three-step process. First, it identifies seed points around which the primary voids will be constructed. This is done by selecting cells from the final output of the Genetic Algorithm (i.e., void block cells) that have a luminosity density lower than the background. VEGA then considers the centers of these cells as seed points for the subsequent steps. Next, it calculates the distance $d_s$ from each seed point to the nearest galaxy. Finally, VEGA sorts the seed points in descending order by this distance, so that those farthest from any galaxy are prioritized first.

\subsection{Final Structure of Voids
\label{sec:mtd_fsv2}}
The final phase consists of three steps that transform the void block cells into a refined catalog of voids. First, VEGA identifies the structure of the primary voids. It begins with the seed point that has the largest distance to its nearest galaxy $d_s$, as determined in the previous phase (section \ref{sec:mtd_fsv}). VEGA then examines all cells whose centers lie within this distance and are part of the void block cells. From this subset, it selects cells that have one or no galaxy within a radial distance $D$ from the cell center, defined as:
\begin{equation}
\label{eq:3.5} 
D = \left|\bar{D}_{nn}^{Gr} - \bar{D}_{nn}^{ga}\right| \,,
\end{equation}
where $\bar{D}_{nn}^{ga}$ and $\bar{D}_{nn}^{Gr}$ are the mean distances to the nearest cell center before and after adding the grid points, respectively. From the cells that satisfy these conditions, VEGA locates the largest contiguous group and labels it as a primary void. This procedure is repeated for the remaining seed points, following the order of decreasing distance.

In the second step, with the set of primary voids identified, VEGA constructs the final structure of voids. It begins by sorting the primary voids based on their radii and selects the largest one as the initial void. For each subsequent primary void, VEGA checks whether it shares two or more common cell faces with any previously labeled void. If no neighbors are found, the void is labeled separately. If one or more neighboring voids exist, VEGA merges the new void with the one that shares the most boundaries, thereby extending the existing structure rather than creating a new label. This process continues until all primary voids have been processed.

In the final step, VEGA excludes any voids that contain marginal cells, which are shown in gray in Figure ~\ref{fig:1}. These voids are incomplete and truncated by the dataset boundaries, making their radii and density contrast measurements unreliable. In addition, VEGA removes all small voids to ensure that minor gaps in the walls or between filaments are not misclassified as voids \cite{hoyle2002voids,sutter2012public}. Prior studies (e.g., \cite{sutter2015vide,schuster2023cosmic}) have shown that the minimum radius cut is strongly dependent on the mean tracer spacing $\bar{n}^{-1/3}$, due to the impact of shot noise. Regarding the dataset described in Section \ref{sec:data}, this threshold corresponds to approximately $5 \ Mpc$, and therefore voids with an effective radius of $R_{void} < 5$ $Mpc$ are excluded. The effective radius of a void is calculated as:
\begin{equation}
\label{eq:3.8}
R_{void} = \sqrt[3]{\frac{3V_{void}}{4\pi}} \,,
\end{equation}
where $V_{void}$ is the total volume of void block cells assigned to that void, as determined in the third phase (Section~\ref{sec:mtd_VT}). VEGA then examines all remaining voids to identify internal holes. If any are detected, it divides the affected void into two separate voids, ensuring they properly surround the enclosed regions. These holes correspond to over-dense regions, combinations of multiple cells that are not part of the void block cells in the final GA result, typically representing small galaxy groups embedded within a surrounding under-dense environment. Consequently, the voids identified by VEGA consist of contiguous cells corresponding to both grid points and galaxies, characterized by a negative luminosity density contrast ($\delta_L < 0$) and an effective radius greater than $5$. These final voids constitute the void catalog, and the galaxies contained within them are classified as void galaxies.

\section{Results and Discussion
\label{sec:res}}
In this section, the results are presented beginning with an investigation into how different values of the Genetic Algorithm (GA) parameters $\epsilon$ and $r_{\mu}$ influence the filtering of background galaxies. This analysis is based on comparisons of luminosity density contrast, background fraction, and the number of generations required under various parameter configurations. Then, the effect of different grid spacings $d_{G}$ is examined, focusing both on the GA filtering outcomes and on the properties of the final voids identified by the VEGA method. These results are compared with those obtained using the Aikio-Mähönen (AM) method \cite{aikio1998simple}, with both methods applied to the same galaxy distribution described in Section~\ref{sec:data}.

The AM method was chosen for comparison due to its fundamentally different void identification approach. This algorithm partitions the dataset space into cells, which are classified as either empty or filled. Filled cells are further categorized into wall and field galaxies based on third-nearest-neighbor distances \cite{hoyle2002voids}. Empty cells are grouped into subvoids using the climbing algorithm \cite{schmidt2001size}, which are subsequently merged into larger voids. Thus, grid points are not used as part of its detection framework, in contrast to VEGA which interprets them as galaxies without luminosity. Voronoi tessellation is also not employed in the AM method, whereas VEGA relies on it to partition the dataset space. Luminosity density plays no role in the void detection process, and the strategy for assembling primary voids into final ones follows a different procedure.

Genetic Algorithm has two parameters, as mentioned in Section~\ref{sec:mtd_GA}: the $\epsilon$ parameter, which regulates the number of chromosomes in each generation, and the $r_{\mu}$ parameter, which is the mutation rate and controls the rate of mutations during generation transitions. By increasing the $\epsilon$ value from 1 to 1000 (i.e., 1, 10, 100, and 1000), while keeping other parameters constant ($r_{\mu}=0.01$ and $d_{G}=5$), the number of chromosomes in each generation increases, leading to an expanded chromosome pool. According to the formula for the number of chromosomes ($N_{ch}$) given in Equation~\ref{eq:3.4}, with $\epsilon=1$, $N_{ch}$ is on the order of $10^{-3}$ of the galaxies in the dataset, whereas with $\epsilon=1000$, $N_{ch}$ is of the same order as the dataset's total number of galaxies.

The final outputs of the Genetic Algorithm analysis (i.e., void block cells) obtained for these $\epsilon$ values are presented in Table~\ref{tab:tab_1}. As seen in the table, increasing $\epsilon$ from 1 to 1000 results in a slight decrease in the average luminosity density contrast ($\delta_{L}$), from $-0.85$ at $\epsilon=1$ to $-0.89$ at $\epsilon=1000$. Regarding the background volume fraction ($V_{b}$), the void block cells contain on average $81\%$ of the background at $\epsilon=1$, compared to $73\%$ at $\epsilon=1000$. Thus, as $\epsilon$ increases, the background percentage becomes smaller and GA filtering becomes more intensive, meaning that more cells are eliminated, retaining fewer cells with lower densities. Additionally, the number of generations ($N_{G}$) that GA progresses through before meeting the halting conditions (Section~\ref{sec:mtd_GA}) varies with the number of chromosomes. As shown in Table~\ref{tab:tab_1}, the number of generations increases on average from $11$ at $\epsilon=1$ to $89$ at $\epsilon=1000$.

\begin{table*}
\centering
\begin{tabular}{c|cccc}
\hline
\hline
$\epsilon$ & $1$ & $10$ & $100$ & $1000$ \\
\hline
$\delta_{L}$ & $-0.85\pm0.002$ & $-0.87\pm0.004$ & $-0.88\pm0.004$ & $-0.89\pm0.003$\\
$V_{b}(\%)$ & $81.4\pm0.4$ & $78.9\pm0.6$ & $75.61\pm0.6$ & $72.75\pm0.7$\\
$N_{G}$ & $11\pm3$ & $41\pm8$ & $69\pm9$ & $89\pm10$\\
\hline
$r_{\mu}$ & $0.01$ & $0.05$ & $0.1$ & $0.5$ \\
\hline
$\delta_{L}$ & $-0.88\pm0.004$ & $-0.86\pm0.003$ & $-0.86\pm0.004$ & $-0.84\pm0.002$\\
$V_{b}(\%)$ & $75.6\pm0.6$ & $79.2\pm0.4$ & $80.1\pm0.4$ & $82.2\pm0.3$\\
$N_{G}$ & $69\pm9$ & $18\pm3$ & $11\pm2$ & $4\pm1$\\
\hline
$d_{G}$ & $5$ & $4$ & $3$ & $No\ Grid$ \\
\hline
$\delta_{L}$ & $-0.88\pm0.004$ & $-0.91\pm0.003$ & $-0.93\pm0.002$ & $-0.8\pm0.007$\\
$V_{b}(\%)$ & $75.6\pm0.6$ & $81.1\pm0.5$ & $87.1\pm0.3$ & $60.8\pm0.9$\\
$N_{G}$ & $69\pm9$ & $68\pm9$ & $68\pm8$ & $71\pm11$\\
\hline
\hline
\end{tabular}
\caption{Void block cell statistics are presented as mean values with standard deviation errors for different Genetic Algorithm parameters $\epsilon$ and $r_{\mu}$, and grid spacings $d_{G}$, based on 100 independent runs for each parameter configuration.}
\label{tab:tab_1}
\end{table*}

Then, by increasing the $r_{\mu}$ parameter value from $0.01$ to $0.5$ ($0.01$, $0.05$, $0.1$, $0.5$), while keeping other parameters constant ($\epsilon=100$ and $d_{G}=5$), the algorithm reduces the suppression of mutations. This occurs because more random numbers generated by the mutation operator fall below the mutation rate, resulting in a greater number of random mutations being applied in each generation transition. This change influences the performance of the GA. Cells that correspond to denser regions, which have lower volume, shorter distance to the nearest-neighbor and higher luminosity density, would normally lower the score of a chromosome and be eliminated in earlier generations. With a higher mutation rate, these cells have a greater chance of being reconsidered by the mutation operator. As a result, the GA tends to favor chromosomes that contain a significant number of cells inside dense regions. Their low nearest‑neighbor distance reduces the score, but their total volume together with their luminosity density can increase it. In this situation, the high rate of random mutations prevents the GA from evolving more suitable chromosomes over multiple generations. The statistics of the void block cells for different mutation rate values are detailed in Table~\ref{tab:tab_1}. As $r_{\mu}$ increases from $0.01$ to $0.5$, the luminosity density contrast ($\delta_{L}$) on average becomes more positive, shifting from $-0.88$ at $r_{\mu}=0.01$ to $-0.84$ at $r_{\mu}=0.5$. In the context of the background volume fraction ($V_{b}$), the fraction increases from $75\%$ to $82\%$, respectively, with the rise in mutation rate. The number of generations ($N_{G}$) in the GA analysis decreases from $69$ generations to $4$ generations for $r_{\mu}=0.01$ and $r_{\mu}=0.5$, respectively.

The impact of two GA parameters on the final void catalog statistics is also examined. When increasing $\epsilon$ from 1 to 1000 while keeping other parameters constant ($d_{G}=5$ and $r_{\mu}=0.01$), the resulting void statistics exhibit no significant changes. The number of voids in the final catalog ($R_{void}\ge5$ $Mpc$) decreases slightly from $734$ at $\epsilon=1$ to $728$ at $\epsilon=1000$, while the mean voids radii increase modestly from $12.32$ to $12.51$. Regarding density contrasts, the mean luminosity density contrast remains constant at $–0.94$ across all runs, whereas the mean number density contrast becomes slightly more negative, changing from $–0.75$ to $–0.76$ with increasing $\epsilon$. These minimal variations in the void statistics indicate that the resulting void catalog is almost insensitive to changes in $\epsilon$, likely due to the large number of galaxies in the input dataset (Section \ref{sec:data}), which according to Eq. \ref{eq:3.4}, yields approximately $270$ chromosomes per generation when $\epsilon=1$, thereby providing sufficient coverage of the GA parameter space. While this pool may influence the statistics of void block cells, the subsequent phases of VEGA mitigate its impact on the final void catalog.

In contrast, varying the mutation rate $r_{\mu}$ from $0.01$ to $0.5$ (with $d_{G}=5$ and $\epsilon=100$ kept constant) produces more pronounced effects. The number of final voids increases from $731$ to $779$, while the mean voids radii decrease from $12.43$ to $11.74$. The mean sphericity improves from $0.64$ to $0.69$, indicating more spherical voids. Both density contrasts become less negative; the mean luminosity density contrast shifts from $–0.94$ to $–0.91$, and the mean number density contrast from $–0.76$ to $–0.71$. These results suggest that high mutation rates lead to voids that are slightly smaller, more spherical, and more densely populated. The observed trends are consistent with the expected influence of elevated mutation rates discussed earlier.

After examining the GA parameters, the effect of different grid spacings ($d_{G}$) on the void block cell statistics is addressed in Table~\ref{tab:tab_1}. Since the minimum radius cut for the voids is set to 5 Mpc (as described in Section \ref{sec:mtd_fsv2}), the maximum grid spacing should be equal to this minimum radius. If a larger grid spacing is used, some small voids may fall between grid points, so the grid‑point utilizing would not be applied to them and its effect would become apparent only for large voids. In such a case, small voids in the catalog would remain unaffected by the use of grid points, while large voids would be affected, leading the algorithm to treat them differently. By using a grid spacing less than or equal to the minimum radius cut, all final voids in the catalogue are affected by the grid‑point utilizing, and therefore the algorithm treats all voids the same in the identification process. Decreasing the $d_{G}$ parameter from $5 \ \mathrm{Mpc}$ to $3 \ \mathrm{Mpc}$, while keeping other parameters constant ($\epsilon=100$ and $r_{\mu}=0.01$), results in a more negative average luminosity density contrast ($\delta_{L}$), shifting from $-0.88$ to $-0.93$. When the algorithm is performed without using grid points, $\delta_{L}$ becomes more positive, averaging around $-0.8$. Regarding the background volume fraction ($V_{b}$), reducing $d_{G}$ leads to a higher fraction, increasing from an average of $75\%$ at $d_{G}=5$ to $87\%$ at $d_{G}=3$, while the no-grid run yields an average $V_{b}$ of $60\%$. The number of GA generations ($N_{G}$) remains nearly unchanged across different grid spacings, averaging between $68$ and $69$ for grid-based runs and reaching $71$ for the run without grid points.

As shown in Table~\ref{tab:tab_1}, the standard deviations of the luminosity density contrast ($\delta_{L}$) and background volume fraction ($V_{b}$) are relatively small compared to their mean values, whereas the number of generations ($N_{G}$) exhibits significantly larger variability. This discrepancy arises from the inherently stochastic nature of the Genetic Algorithm. The data in Table~\ref{tab:tab_1} represents results from 100 independent runs per parameter configuration, and the elevated error in $N_{G}$ stems from the randomized initial population in each run. Although the Genetic Algorithm begins with different chromosome pool in each run, it consistently converges to highly similar outcomes for $\delta_{L}$ and $V_{b}$. This consistency suggests that for each parameter setup, the GA tends to navigate toward a top-performing region of the parameter space. While the specific stopping point varies slightly between runs due to randomness, the algorithm reliably halts in a zone closely aligned with those from other trials, indicating convergence toward robust and reproducible solutions. Notably, the final values of $\delta_{L}$ and $V_{b}$ across runs differ by less than $1\%$, underscoring the stability and reliability of the optimization process.

\begin{figure*}[t]
\centering
\includegraphics[width=1\textwidth]{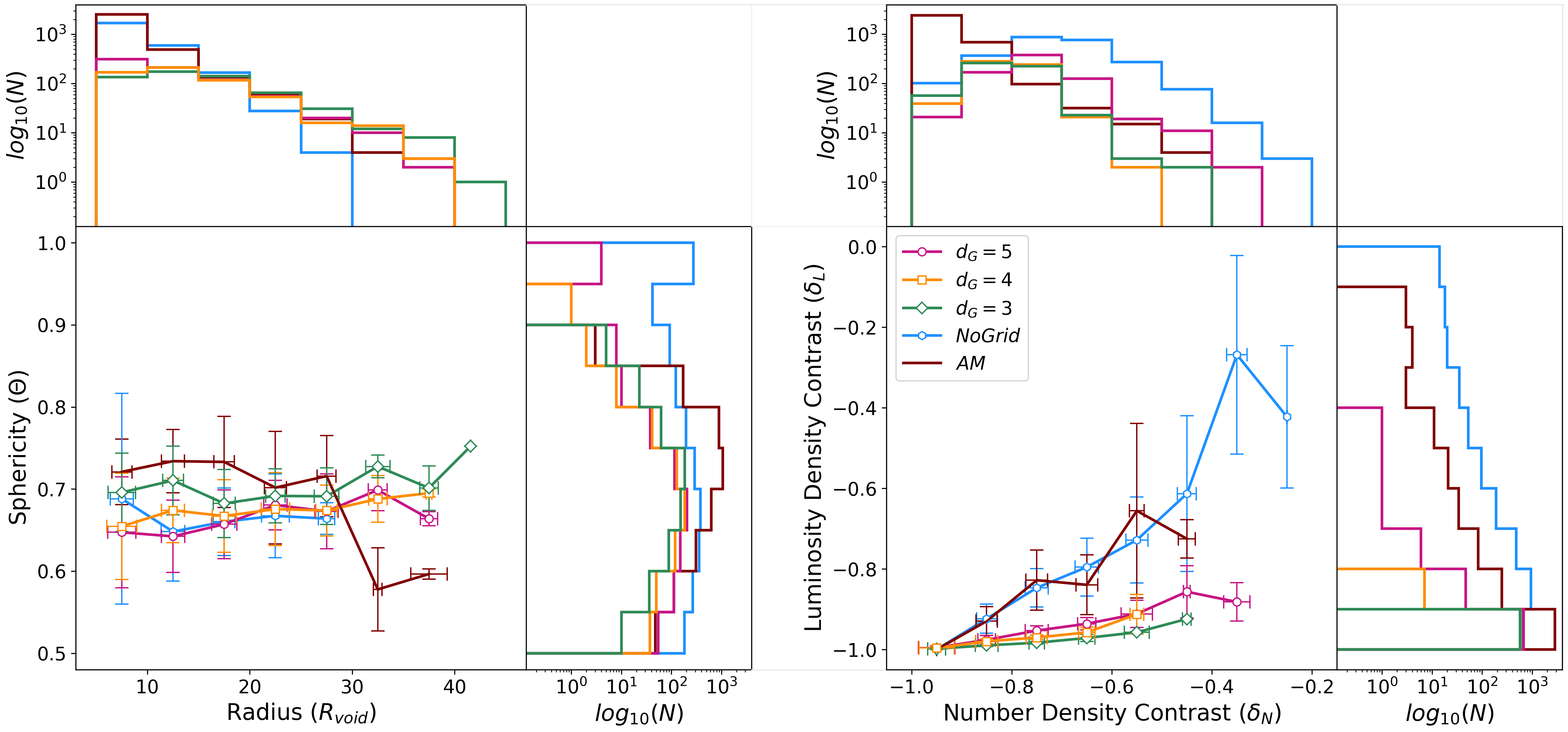}
\caption{
\textit{(Left Panel)} displays median trends and associated errors for void sphericity ($\Theta$) versus radius ($R_{\mathrm{void}}$). \textit{(Right Panel)} shows the median trends and errors between luminosity density contrast ($\delta_{L}$) and number density contrast ($\delta_{N}$). Histograms of these parameters are displayed alongside the median plots. Results are presented for VEGA runs with varying grid spacings $d_{G}$, alongside the results obtained using the AM method.}
\label{fig:3}
\end{figure*}

At the final step, the performance and the voids identified by VEGA are compared with those identified by the AM method. For the VEGA runs, the parameters $\epsilon$ and $r_{\mu}$ are set to 100 and 0.01, respectively. This choice was due to the regulation of the GA filtering for the runs using grid points in such a way that the remaining background (i.e., void block cells) has a volume fraction in the range of $70\%$ to $90\%$, as mentioned in previous studies (\cite{tavasoli2013challenge, hellwing2021caught, curtis2024properties, mansour2025j}), and this configuration was also set for the run without using grid points to allow reliable comparison of the results. The results and the final void characteristics are presented in Figure~\ref{fig:3} and Table~\ref{tab:tab_2}.

\begin{table*}
\centering
\begin{tabular}{c|ccccc}
\hline
\hline
\multirow{2}{*}{Parameter} & \multicolumn{4}{c}{VEGA} & \multirow{2}{*}{AM} \\
\cline{2-5}
 & $d_{G}=3$ & $d_{G}=4$ & $d_{G}=5$ & $No\ Grid$ & \\
\hline
$N_{voids}$ & 571 & 585 & 731 & 2485 & 3264 \\
$R_{void}$ & $15.17\pm6.77$ & $13.89\pm6.26$ & $12.43\pm6.11$ & $9.12\pm3.66$ & $8.44\pm3.95$ \\
$\Theta$ & $0.69\pm0.06$ & $0.66\pm0.07$ & $0.64\pm0.09$ & $0.68\pm0.17$ & $0.71\pm0.07$ \\
$\delta_{L}$ & $-0.98\pm0.01$ & $-0.97\pm0.02$ & $-0.94\pm0.04$ & $-0.78\pm0.23$ & $-0.95\pm0.09$ \\
$\delta_{N}$ & $-0.82\pm0.07$ & $-0.81\pm0.06$ & $-0.76\pm0.09$ & $-0.69\pm0.12$ & $-0.93\pm0.08$ \\
\hline
\hline
\end{tabular}
\caption{Mean values and errors (standard deviations) of void characteristics—number of voids ($N_{\mathrm{voids}}$), void radius ($R_{\mathrm{void}}$), sphericity ($\Theta$), luminosity density contrast ($\delta_{L}$) and number density contrast ($\delta_{N}$)—identified using VEGA with varying grid spacings ($d_{G}$) and the AM method.}
\label{tab:tab_2}
\end{table*}

The presented parameters are: luminosity density contrast ($\delta_{L}$), which is measured by using the total luminosity of the galaxies inside each void, the total volume of all the cells that comprise each void (both corresponding to grid points and galaxies), and the background luminosity density as discussed in Section~\ref{sec:mtd_GA} (Equation~\ref{eq:3.2}); number density contrast ($\delta_{N}$), measured similarly to $\delta_{L}$ but using the number of galaxies inside each void instead of their luminosity; and the sphericity parameter ($\Theta$), which is calculated as the ratio of the volume of the void contained within a sphere of radius $R_{void}$ divided by the total volume of the void \cite{tavasoli2013challenge}. Thus, completely spherical voids have a sphericity of $\Theta=1$, while a lower $\Theta$ indicates a void that is less spherical in shape. From these results, the effect of using grid points with different grid spacings ($d_{G}$) can be seen. Due to the effect of different $d_{G}$ values discussed earlier in this section (Table~\ref{tab:tab_1}), reducing the grid spacing causes the total background volume of the void block cells to increase. This is due to the higher number of grid points that can capture empty spaces in the dataset more effectively, and also the void boundaries get closer to structures like clusters, groups, and filaments (as depicted in Figure~\ref{fig:2}). As the background volume of the void block cells gets larger, the primary voids gain more volume and their common sides with neighboring primary voids become larger. This leads to more grouping of primary voids into final voids, resulting in larger final voids with a lower overall count. This increase in void volume results in more negative luminosity ($\delta_{L}$) and number density contrasts ($\delta_{N}$), and slightly more spherical final voids. The variation in the number of final voids indicates that, for the dataset used in this study (Section \ref{sec:data}), a grid spacing ($d_{G}$) of about 4 Mpc is more stable than 5 Mpc and 3 Mpc, since the reduction from 5 Mpc to 4 Mpc produces a substantial change in void count, whereas the difference between 4 Mpc and 3 Mpc is negligible.

In the run without using grid points, the results show on average smaller voids with more positive luminosity ($\delta_{L}$) and number density contrasts ($\delta_{N}$), and average sphericity in a similar range. As can be seen in Figure~\ref{fig:3} and Tables~\ref{tab:tab_1} and \ref{tab:tab_2}, using grid points brings advantages to the voids identified by VEGA in comparison to runs without using grid points. With grid points, while the mean sphericity of voids remains nearly in the same range, the density contrasts with respect to the background ($\delta_{L}$ and $\delta_{N}$) become more negative. Also, the errors of the parameters are relatively high compared to the mean values in runs without grid points, but these errors are much smaller when grid points are used. The voids identified by AM are also presented so that the performance of VEGA in different configurations can be compared to another algorithm with a completely different void identification procedure.

\section{Summary
\label{sec:sum}}

In this study, a new void identification method based on Genetic Algorithm analysis (VEGA), is introduced. The VEGA approach relies on the Voronoi tessellation of the input distribution and incorporates grid points into the dataset, treating them as galaxies with zero luminosity. This addition alters the shape of the Voronoi cells, allowing VEGA to more effectively capture empty regions and produce voids with smoother boundaries, reducing the occurrence of sharp-edged or irregularly shaped cells, as discussed in Section \ref{sec:mtd_grd}. In addition to incorporating grid points, VEGA applies a Genetic Algorithm-based filtering process to eliminate over-dense regions while preserving continuous combination of cells, referred to as void block cells (Section \ref{sec:mtd_GA}). These cells form the basis for subsequent phases of void identification. For this filtering, VEGA employs luminosity density contrast instead of the traditional number density contrast, allowing for a more accurate representation of over-densities in the luminosity field. Furthermore, VEGA uses an adaptable background luminosity density to account for the influence of grid points on both the Voronoi cells and the background luminosity density, thereby removing the misclassification of over-densities due to the effects introduced by grid points (Section \ref{sec:mtd_LD}).

Employing the galaxy distribution from the Millennium simulation as described in Section \ref{sec:data}, the effects of varying GA parameters and grid spacings on both void block cell statistics and the final void catalog were probed in Section \ref{sec:res}. These results were then compared to voids identified by the AM method applied to the same dataset, providing a basis for evaluating VEGA against a fundamentally different void identification approach. Nevertheless, it remains challenging to fully evaluate the success of any void identification approach due to the absence of a clear consensus on the definition of these vast under-dense structures, as different studies have adopted diverse methods and criteria for defining cosmic voids. The results implied that the voids identified by VEGA are generally consistent with the definition of cosmic voids as continuous regions exhibiting extreme under-density and occupying a significant fraction of the dataset’s volume. They also align with voids identified in previous studies that employed alternative identification methods (e.g., \cite{neyrinck2008zobov,sutter2015vide,nadathur2019revolver}). These findings reinforce the reliability of the VEGA method in identifying and characterizing cosmic voids.

Further studies could explore how characteristics of the input distribution influence the performance and outcomes of the method, using datasets with larger box sizes and higher tracer counts than the one used in this work, which primarily served to introduce and demonstrate this new void-finding approach. Additional investigations are needed to apply VEGA to datasets from different cosmological simulations based on the standard model (e.g., \cite{nelson2019illustristng,ayromlou2021galaxy}), as well as to observational survey data (e.g., \cite{almeida2023eighteenth,adame2024early}) with varying background densities. Moreover, further steps towards developing a version suitable for void identification in cosmological observations should be considered to enhance its applicability to real-world data, while comparative studies involving other void identification techniques (e.g., \cite{rincon2025desivast}) could help clarify VEGA’s performance across diverse contexts.

\acknowledgments
We would like to express our gratitude to Shahram Khosravi for his valuable support throughout this project. We also thank Ghazaleh Mahjoub for the insightful conversations. Our sincere appreciation goes to the anonymous referee, whose thoughtful evaluation and constructive critique significantly enhanced the clarity, depth, and scholarly rigor of this work.

\bibliographystyle{JHEP}
\bibliography{VEGA}

\end{document}